\documentclass[11pt, a4paper]{article}
\usepackage{jcappub}
\usepackage{bm}
\usepackage{graphicx}
\usepackage{epstopdf}
\usepackage[all]{nowidow}

\newcommand{\MP}{M_{\rm Pl}}

\newcommand{\df}{\dot{\phi}}
\newcommand{\ddf}{\ddot{\phi}}
\newcommand{\ee}{\epsilon_{\rm end}}
\newcommand{\pe}{\phi_{\rm end}}
\newcommand{\Qe}{Q_{\rm end}}

\newcommand{\PkO}{A_s}
\newcommand{\Pk}{\Delta^2_{\mathcal{R}}(k)}

\title{Observational Constraints on\\ Monomial Warm Inflation}

\author{Luca Visinelli}
\affiliation{Nordita, KTH Royal Institute of Technology and Stockholm University,\\ SE-106 91 Stockholm, Sweden}
\emailAdd{Luca.Visinelli@studio.unibo.it}

\abstract{Warm inflation is, as of today, one of the best motivated mechanisms for explaining an early inflationary period. In this paper, we derive and analyze the current bounds on warm inflation with a monomial potential $U\propto \phi^p$, using the constraints from the PLANCK mission. In particular, we discuss the parameter space of the tensor-to-scalar ratio $r$ and the potential coupling $\lambda$ of the monomial warm inflation in terms of the number of e-folds. We obtain that the theoretical tensor-to-scalar ratio $r\sim 10^{-8}$ is much smaller than the current observational constrain $r \lesssim 0.12$, despite a relatively large value of the field excursion $\Delta \phi \sim 0.1\MP$. Warm inflation thus eludes the Lyth bound set on the tensor-to-scalar ratio by the field excursion.}

\arxivnumber{}

\begin{document}
\maketitle

\section{Introduction}

Inflation~\cite{kazanas1980, starobinsky1980, guth1981, sato1981, mukhanov1981, albrecht1982a, linde1982, linde1983} provides us with a possible explanation for a number of severe problems posed in the standard Big-Bang cosmology like the observed flatness, homogeneity, and the lack of relic monopoles~\cite{mukhanov_rev, linde_book, kolb_book, bergstrom_book, weinberg_book}. This is achieved by means of a microphysical model, in which one dynamical field (the ``inflaton'') evolves under the influence of a nearly-flat potential, providing an approximately constant expansion rate. Inflation also serves as a mechanism to predict the acoustic peaks in the Cosmic Microwave Background Radiation (CMBR)~\cite{bennett2003, bennett2013} and generate the perturbations observed in the CMBR from the evolution of primordial quantum vacuum fluctuations during inflation~\cite{guth1982, hawking1982, starobinsky1982, bardeen1983, steinhardt1984}.

A different possible source to generate perturbations in agreement with observations are thermal fluctuations arising from a dissipation term~\cite{abbott1982, albrecht1982b, morikawa1984, hosoya1984, moss1985, lonsdale1987, yokoyama1988, liddle1989}. This alternative mechanism has been considered in warm inflation~(\cite{berera1995a, berera1995b}, see also the review in Ref.~\cite{berera2009}) which is, as of today, one of the best motivated inflation models to achieve an early inflationary period. In fact, the warm inflation model with a quartic potential is consistent with the inner confidence region of the PLANCK mission data~\cite{bartrum2013, ramos2013, basterogil2011, basterogil2013, basterogil2014}. In warm inflation models, the energy density in the inflaton field converts into a non-negligible population of relativistic degrees of freedom (from here on ``radiation'') at a rate $\Gamma$, which is sufficiently strong compared to the Hubble rate $H$~\cite{berera2000,taylor2000,hall2004}. In the simplest possible model, radiation is close to thermal equilibrium, and both the particle production rate and the dissipation rate of the inflaton field depend on $\Gamma$ only~\cite{gleiser1994, berera1998, yokoyama1999}. Non-gaussianities in the thermal spectrum have been used to constrain the dependence on the temperature of the dissipation rate $\Gamma$~\cite{moss2007, moss2011}. Recent approaches deal with the behavior of a thermal fluctuation in an expanding Universe, on the basis of the equivalence principle~\cite{berera2005, delcampo2007, moss2008, graham2009, visinelli2015}. Successful warm inflation models have been built within supersymmetric theories, in which a decay mechanism is implemented with the inflaton decaying into light radiation fields through a heavy particle intermediary~\cite{berera2003,berera2005,basterogil2005a, basterogil2005b, moss2006, matsuda2010, basterogil2016}.

In the era of precision cosmology, inflation theories have to be tested against the robust measurements obtained by the Planck mission~\cite{planck}. For warm inflation, these tasks have been addressed in Ref.~\cite{oliveira2001, oliveira2002} against measurements from the COBE satellite~\cite{smoot1992}, using the power spectrum and the scalar spectral tilt previously derived~\cite{berera2000,taylor2000}. The parameter space of warm inflation with a $\phi^4$ potential has been discussed in Ref.~\cite{bartrum2013}, while the constraints with various potentials from ``chaotic'' inflation~\cite{linde1983} have been given in Refs.~\cite{ramos2013, basterogil2011, basterogil2013, basterogil2014}.  A throughout analysis of the constraints in warm inflation with an axion-like inflaton particle has also been considered in the so-called ``natural warm inflation'' scenario~\cite{visinelli2011}. The observational bounds on the warm inflation version of the Dirac-Born-Infeld model have been discussed in Ref.~\cite{cai2011}.


In this paper, we investigate the current bounds on the parameters coming from the recent measurements from the Planck satellite, and we set the order of magnitude of the important quantities appearing in warm inflation. In particular, we use the measurements of the density power spectrum and its tilt at a specific pivot scale, from which the Hubble rate $H$, the dissipation rate $\Gamma$, and the tensor-to-scalar ratio $r$ depend. When a monomial potential is introduced, we obtain the results summarized in Table~\ref{Table_magnitude}. In the following we refer to warm inflation with a $U\propto \phi^p$ potential as a ``monomial warm inflation'' theory. We also discuss on the magnitude of the field excursion $\Delta \phi$, which discerns between small- and large-field inflation models depending on the magnitude of the excursion when compared with the reduced Planck mass $\MP$. We obtain that monomial warm inflation is a ``middle''-field model, in which $\Delta \phi \sim 0.1\,\MP$, except for a region in the parameter space in which we obtain a small field theory with $\Delta \phi \ll \MP$.

This paper is organized as follows. We first review the main equations describing warm inflation in the slow-roll regime in Sec.~\ref{The warm inflation scenario}, and the expressions for the cosmological observables in Sec.~\ref{Perturbations from inflation}. In Sec.~\ref{Shape of the potential and slow-roll parameters}, we restrict our discussion to that of monomial potentials, while in Sec.~\ref{Number of e-folds} we analyze the dynamic of the inflaton field using the slow-roll conditions and number of e-folds for sufficient inflation. We discuss results in the strong limit of warm inflation in Sec.~\ref{Strong limit of warm inflation}, and we show how observation constraints the parameter space of monomial warm inflation in Sec.~\ref{application}.

\section{Slow-roll regime of warm inflation}\label{The warm inflation scenario}

In the warm inflation scenario, the inflaton field appreciably converts into radiation during the inflationary period. This mechanism is parametrized by the appearance of the dissipative term $\Gamma$ in the dynamics of the inflaton field. In the following we assume~\cite{berera1995a, berera1995b} that radiation thermalizes on a time scale much shorter than $1/\Gamma$, so that the energy density in radiation is
\begin{equation}\label{definition_radiation}
\rho_r = \alpha\,T^4,\quad\hbox{with} \quad \alpha = \frac{\pi^2}{30}\,g_*(T),
\end{equation}
where $g_*(T)$ is the number of relativistic degrees of freedom of radiation at temperature~$T$. Here we do not specify $g_*(T)$ in the equations, but in the results obtained in Sec.~\ref{Warm monomial inflation} we will always use $g_*(T) = 228.75$, corresponding to the number of relativistic degrees of freedom in the minimal supersymmetric Standard Model at temperatures $T$ greater than the electroweak phase transition.

Warm inflation is achieved when thermal fluctuations dominate over quantum fluctuations, or~\cite{berera1995a, berera1995b}
\begin{equation} \label{condition_warm_inflation}
H(T) < T,
\end{equation}
where $H$ is the Hubble expansion rate at temperature $T$. The effectiveness at which the inflaton converts into radiation is measured by the ratio
\begin{equation} \label{effectiveness}
Q = \frac{\Gamma}{3H},
\end{equation}
where the limit $Q \ll 1$ ($Q \gg 1$) represents the weak (strong) regime of warm inflation.

In the following, we model the inflaton field with a scalar field $\phi = \phi(x)$ moving in a potential $U = U(\phi)$. The evolution of the inflaton field in a Friedmann-Robertson-Walker metric is described by~\cite{kolb_book}
\begin{equation} \label{eq_motion}
\ddot{\phi} + (3H+\Gamma)\dot{\phi} + U_\phi = 0,
\end{equation}
where a dot indicates the derivation with respect to the cosmic time $t$ and $U_\phi = \partial U/\partial\phi$. The conservation of the total energy of the system requires that the energy density of radiation satisfies
\begin{equation} \label{energy_conservation_radiation}
\dot{\rho_r} + 4H\rho_r = \Gamma\,\dot{\phi}^2,
\end{equation}
with the term on the RHS of Eq.~(\ref{energy_conservation_radiation}) describing the effectiveness of conversion of the inflaton field into radiation. The Friedmann equation reads
\begin{equation} \label{friedmann}
H^2 = \frac{\rho}{3\MP^2} = \frac{1}{3\MP^2}\left(\frac{1}{2}\dot{\phi}^2+U+\rho_r\right),
\end{equation}
where $\MP = 2.435 \times 10^{18}{\rm~GeV}$ is the reduced Planck mass and the term within brackets describes the total energy density of the system.

Inflation takes place when the potential $U$ is approximately flat and the potential energy dominates over all other forms of energy, assuring an approximately constant value of the Hubble expansion rate. During this period, which is known as the slow-roll regime of the inflaton field, higher derivatives in Eqs.~(\ref{eq_motion}) and~(\ref{energy_conservation_radiation}) can be neglected,
\begin{equation}\label{slow_roll_conditions}
\ddot{\phi} \ll H \,\dot{\phi},\quad\hbox{and}\quad \dot{\rho}_r \ll H\,\rho_r.
\end{equation}
In this regime, Eqs.~(\ref{eq_motion}),~(\ref{energy_conservation_radiation}) and~(\ref{friedmann}) read
\begin{eqnarray}
\dot{\phi} &\simeq& -\frac{U_\phi}{3H+\Gamma}, \label{eq_motion_slow_roll}\\
\rho_r &\simeq& \frac{3Q}{4}\,\dot{\phi}^2, \label{energy_conservation_sl} \\
H^2 &\simeq& \frac{1}{3\MP^2}\,U, \label{friedmann_sl}
\end{eqnarray}
where here and in the following we use the symbol ``$\simeq$'' for an equality that holds in the slow-roll regime. From Eq.~\eqref{friedmann_sl} it follows that a shallow potential $U$ gives rise to a nearly constant expansion rate $H$.

The slow-roll regime can be parametrized by a set of slow-roll parameters $\epsilon$, $\eta$ and $\beta$, defined by~\cite{liddle1994}
\begin{equation}\label{slow_roll_parameters}
\epsilon = \frac{\MP^2}{2}\,\left(\frac{U_\phi}{U}\right)^2, \quad \eta = \MP^2\,\frac{U_{\phi\phi}}{U},\quad\beta = \MP^2\,\left(\frac{\Gamma_{\phi}\,U_\phi}{\Gamma\,U}\right).
\end{equation}
From the definition for $\beta$ in Eq.~\eqref{slow_roll_parameters} it can be shown that
\begin{equation}
\frac{\dot{\Gamma}}{H\,\Gamma} \simeq -\frac{\beta}{1+Q}.
\end{equation}
To assure the conditions expressed in Eq.~(\ref{slow_roll_conditions}), we first differentiate Eqs.~(\ref{eq_motion_slow_roll}),~(\ref{energy_conservation_sl}) and~(\ref{friedmann_sl}), obtaining
\begin{eqnarray}
\frac{\dot{H}}{H^2} &\simeq& -\frac{\epsilon}{1+Q}, \label{relation_epsilon_Hdot}\\
\frac{\ddot{\phi}}{H\,\dot{\phi}} &\simeq& -\frac{1}{1+Q}\left(\eta-\beta+\frac{\beta-\epsilon}{1+Q}\right),\\
\frac{\dot{\rho}_r}{H\,\rho_r} &\simeq& -\frac{1}{1+Q}\left(2\eta-\beta-\epsilon+2\frac{\beta-\epsilon}{1+Q}\right) \label{relation_parameters_Rdot}.
\end{eqnarray}
The conditions in Eq.~(\ref{slow_roll_conditions}) are met by demanding that~\cite{taylor2000, hall2004, basterogil2005b, moss2008, visinelli2011}
\begin{equation} \label{slow_roll}
\epsilon \ll 1+Q,\quad |\eta| \ll 1+Q,\quad |\beta| \ll 1+Q.
\end{equation}
Eq.~(\ref{slow_roll}) is a generalization of the slow-roll conditions in the cold inflation that takes into account the parameter $Q$; when $Q \ll 1$ the dissipation term can be neglected and the slow-roll conditions reduce to the usual requirements in the cold inflation.

A relationship between $T$ and $U$ valid during the slow-roll regime is obtained by inserting the definition of $\rho_r$ given in Eq.~\eqref{definition_radiation} into Eq.~\eqref{energy_conservation_sl}, and using the expression for $\df$ and $H^2$ in Eqs.~\eqref{eq_motion_slow_roll} and~\eqref{friedmann_sl},
\begin{equation}\label{relation_T_dotphi}
T \simeq \sqrt[4]{\frac{U}{2\alpha}\,\frac{Q_k}{(1+Q_k)^2}\,\frac{\MP^2}{2}\,\left(\frac{U_\phi}{U}\right)^2} = \sqrt[4]{\frac{U}{2\alpha}\,\frac{Q_k}{(1+Q_k)^2}\,\epsilon_k},
\end{equation}
where in the last expression we have inserted $\epsilon_k$ thanks to Eq.~\eqref{slow_roll_parameters}.

\section{Review of the perturbations spectra} \label{Perturbations from inflation}

\subsection{Scalar power spectrum} \label{Scalar power spectrum}

A standard paradigm of the inflationary mechanism states that scalar and tensor fluctuations which emerged during the inflationary epoch later evolved into primordial perturbations of the density profile and gravitational waves, leaving an imprint in the CMBR anisotropy and on large scale structures~\cite{mukhanov1981, guth1982, hawking1982, starobinsky1982, bardeen1983}. It is thus of primarily importance to review the expressions of the observables in warm inflation, to compare the prediction of the theory with measurements.

The spectrum of the adiabatic density perturbations generated during inflation is expressed by the function~\cite{kosowsky1995, leach2003, liddle2003}
\begin{equation} \label{curvature_perturbations}
\Pk \equiv \frac{k^3\,P_{\mathcal{R}}(k)}{2\pi^2} = \PkO\,\left(\frac{k}{k_0}\right)^{n_s(k)-1},
\end{equation}
which depends mildly on the co-moving wavenumber $k$ according to a spectral index $n_s(k)$, discussed later in Sec.~\ref{Scalar spectral index}. Here, we use the notation in Ref.~\cite{planck} and we set $\PkO = \Delta_{\mathcal{R}}^2(k_0)$ at the reference scale $k_0$. The function $\Delta_{\mathcal{R}}^2(k)$ describes the contribution to the total variance of primordial curvature perturbations at a given scale per logarithmic interval in $k$ \cite{bennett2013}. The curvature perturbation spectrum has been measured by the PLANCK mission at 68\%~Confidence Level (CL) at the fixed wave number $k_0 = 0.05{\rm~Mpc}^{-1}$ as~\cite{planck}
\begin{equation} \label{constraint_power_spectrum}
\PkO = \left(2.215^{+0.032}_{-0.079}\right)\times 10^{-9}.
\end{equation}
The WMAP collaboration~\cite{bennett2013} previously reported a similar result $\PkO=(2.445 \pm 0.096)\times 10^{-9}$ at $k_0=0.002\, {\rm Mpc}^{-1}$ with 68\%~CL.

The RHS of Eq.~(\ref{curvature_perturbations}) is evaluated when a given co-moving wavelength crosses outside the Hubble radius during inflation, and the LHS when the same wavelength re-enters the horizon. In Eq.~(\ref{curvature_perturbations}) we have used the notation in Ref.~\cite{bennett2013} for the density perturbations. On a theoretical ground, the shape of the scalar power spectrum is given by
\begin{equation}\label{scalar_spectrum}
\PkO = \left(\frac{H}{\dot{\phi}}\,\langle\delta \phi\rangle\right)^2,
\end{equation}
where $\dot{\phi} \simeq -U_\phi/(3H+\Gamma)$ and $\langle\delta \phi\rangle$ describes the spectrum of fluctuations in the inflaton field. The LHS of Eq.~(\ref{scalar_spectrum}) is computed at the time at which the largest density perturbations on observable scales are produced. To keep track of this, in the following we add a subscript $k$ to the quantities which are evaluated at the horizon crossing. To compute the value of the Fourier transform of the field fluctuations $\delta \phi_q$, a stochastic field approach is often used. The interaction between the inflaton field and radiation can be analysed within the Schwinger-Keldysh approach to non-equilibrium field theory~\cite{schwinger1961, keldysh1964, rammer1986}. In flat spacetime, the system is described by two coupled Langevin equations~\cite{calzetta1988}, where a Gaussian white noise source $\xi_q$ describing the effect of thermal fluctuations appears. The expression describing the evolution of fluctuations in an expanding Universe is obtained by applying the equivalence principle to the non-expanding result, and reads~\cite{berera1995a, berera2005, delcampo2007, moss2008, graham2009, bartrum2013, visinelli2015},
\begin{equation} \label{eq_motion_perturbation}
\delta\ddot{\phi}_q + (3H+\Gamma)\,\delta\dot{\phi}_q + \left(\frac{q^2}{a^2} + U_{\phi\phi}\right)\delta \phi_q = \sqrt{2\,\Gamma\,T}\,a^{-3/2}\,\xi_q.
\end{equation}
The theoretical power spectrum reads~\cite{bartrum2013, ramos2013}
\begin{equation} \label{theoretical_spectrum_full}
\PkO = \left(\frac{H_k^4}{4\pi^2\,\df_k^2}\right)\,\left(1+2\nu_k + \omega_k\right),
\end{equation}
where the term in the first brackets in Eq.~\eqref{theoretical_spectrum_full} corresponds to the power spectrum in the cold inflation model, while the terms $\nu_k$ and $\omega_k$ provide the corrections respectively from the non-trivial occupation numbers of the inflaton and from thermal effects, with
\begin{equation}
\omega_k = \left(\frac{T}{H}\right)\frac{2\sqrt{3}\,\pi\,Q_k}{\sqrt{3+4\pi\,Q_k}}.
\end{equation}
The power spectrum for the cold inflation model is found when both $\nu_k \to 0$ and $\omega_k \to 0$. For $T \gg H$, thermal fluctuations dominate over quantum flucutations and warm inflation applies. In this limit, we obtain the behavior
\begin{equation} \label{omega_limits}
\omega_k \sim \begin{cases} 
T\,\sqrt{\frac{\pi\,\Gamma}{H^3}}, &\hbox{for $Q_k \gg 1$},\\
\frac{2\pi\,\Gamma\,T}{3H^2}, &\hbox{for $Q_k \ll 1$}.
\end{cases}
\end{equation} 
The first expression in Eq.~\eqref{omega_limits} corresponds to the behavior in the strong limit of warm inflation~\cite{taylor2000, hall2004}, while the second expression corresponds to the behavior in the weak limit of warm inflation~\cite{berera1995a}.

\subsection{Scalar spectral index} \label{Scalar spectral index}

The scalar spectral tilt can be defined using Eq.~\eqref{curvature_perturbations} as
\begin{equation} \label{spectral_tilt}
n_s - 1 \equiv \frac{d \ln \Pk}{d \ln k}\bigg|_{k=k_0}.
\end{equation}
The PLANCK mission~\cite{planck} measures the spectral tilt at $k_0 = 0.05{\rm~Mpc}^{-1}$ at 68\%CL as
\begin{equation} \label{spectral_tilt_measured}
n_s = 0.9655 \pm 0.0062.
\end{equation}
The spectral tilt is computed theoretically by taking the logarithmic derivative of the terms in Eq.~\eqref{theoretical_spectrum_full} with respect to $\ln k$, as given in Eq.~\eqref{spectral_tilt}. In the limit where the production of inflaton particles is not efficient $\nu_k \ll1$, we obtain~\cite{berera2000, taylor2000, hall2004}
\begin{equation} \label{spectral_tilt1}
1-n_s = 4\frac{\dot{H}}{H^2} - 2\frac{\ddf}{H_k\,\df}-\frac{\dot{\omega}_k}{H(1+\omega_k)}.
\end{equation}
Taking the derivative of the parameter $\omega_k$, we obtain
$$\frac{\dot{\omega}_k}{H\,\omega_k} = \frac{1}{4}\frac{\dot{\rho}_r}{H\rho_r} - \frac{\dot{H}}{H^2} + \frac{3+2\pi\,Q_k}{3+4\pi\,Q_k}\,\frac{\dot{Q}_k}{H\,Q_k} = \frac{1}{4}\frac{\dot{\rho}_r}{H\rho_r} - \frac{\dot{H}}{H^2} + \left[\frac{1}{2} + \frac{3}{2(3+4\pi\,Q_k)}\right]\,\frac{\dot{Q}_k}{H\,Q_k},$$
or, using the expressions for the derivatives in Eqs.~\eqref{relation_epsilon_Hdot} and~\eqref{relation_parameters_Rdot} plus the identity
\begin{equation}
\frac{\dot{Q}_k}{H\,Q_k} = \frac{\dot{\Gamma}}{H\Gamma} - \frac{\dot{H}}{H^2} = -\frac{1}{1+Q_k}\,\left(\beta_k-\epsilon_k\right),
\end{equation}
we obtain the expression for $\dot{\omega}_k$ as
$$\dot{\omega}_k \simeq -\frac{H\,\omega_k}{1+Q}\,\left[\frac{2\eta_k + \beta_k - 7\epsilon_k}{4} + \frac{6+(3+4\pi)\,Q_k}{(1+Q_k)(3+4\pi Q_k)}\,(\beta_k-\epsilon_k)\right].$$
Eventually, the scalar spectral tilt is given by
\begin{equation} \label{spectral_tilt_warm}
n_s\!-\!1 \!\simeq\! \frac{1}{1 \!+\! Q_k}\,\left[-4\epsilon_k\!+\! 2\left(\eta_k\!-\!\beta_k\!+\!\frac{\beta_k\!-\!\epsilon_k}{1\!+\!Q_k}\right)-\frac{\omega_k}{1\!+\!\omega_k}\left(\frac{2\eta_k \!+\! \beta_k \!-\! 7\epsilon_k}{4}\!+\!\frac{6\!+\!(3\!+\!4\pi)\,Q_k}{(1\!+\!Q_k)(3\!+\!4\pi Q_k)}\,(\beta_k-\epsilon_k)\right)\right].
\end{equation}
The cold inflation regime is achieved when $Q_k \to 0$ and $T \ll H$, or $\omega_k \ll 1$. In this limit, we obtain the known result $n_s -1 = -6\epsilon_k + 2\eta_k$. In warm inflation $T > H$, two different regimes have been considered.

For strong dissipation $Q_k \gg 1$, the terms inside the square brackets containing a $1/(1+Q_k)$ factor can be ignored, so Eq.~\eqref{spectral_tilt_warm} reads
\begin{equation} \label{spectral_tilt_warm_strong}
n_s\!-\!1 \!\simeq\! \frac{1}{Q_k}\,\left[-4\epsilon_k\!+\! 2\left(\eta_k\!-\!\beta_k\!\right)-\frac{\omega_k}{1\!+\!\omega_k}\left(\frac{2\eta_k \!+\! \beta_k \!-\! 7\epsilon_k}{4}\right)\right].
\end{equation}
When $Q_k \gg 1$, we also obtain $\omega_k \gg 1$, so that Eq.~\eqref{spectral_tilt_warm_strong} reads
\begin{equation} \label{spectral_tilt_warm_strong1}
n_s\!-\!1 \!\simeq\! \frac{1}{Q_k}\,\left(-\frac{9}{4}\epsilon_k+\frac{3}{2}\eta_k-\frac{9}{4}\beta_k\right),\quad\hbox{for $Q_k \gg 1$}.
\end{equation}
This latter expression coincides with previous results obtained in the literature~\cite{hall2004}.

For weak dissipation $Q_k \ll 1$, Eq.~\eqref{spectral_tilt_warm} simplifies as
\begin{equation} \label{spectral_tilt_warm_weak}
n_s\!-\!1 \!\simeq\! -6\epsilon_k\!+\! 2\eta_k + \frac{\omega_k}{1\!+\!\omega_k}\left(\frac{15\epsilon_k-2\eta_k - 9 \beta_k}{4}\right).
\end{equation}
Taking the further assumption that $\omega_k \ll 1$, we obtain
\begin{equation} \label{spectral_tilt_warm_weak1}
n_s\!-\!1 \!\simeq\! -6\epsilon_k\!+\! 2\eta_k + 2\pi\,\frac{\Gamma\,T}{3H^2}\,\left(\frac{15\epsilon_k-2\eta_k - 9 \beta_k}{4}\right),\quad\hbox{for $Q_k \ll 1$}.
\end{equation}

\subsection{Tensor power spectrum}

For the primordial tensor fluctuations, the corresponding power spectrum takes the form
\begin{equation}
\Delta_{\mathcal{T}}^2(k) = \Delta_{\mathcal{T}}^2(k_0)\,\left(\frac{k}{k_0}\right)^{n_{\mathcal{T}}},
\end{equation}
where $n_{\mathcal{T}} = -2\epsilon_k$ is the tensor spectral tilt and the power spectrum at the pivotal scale $k_0$ is
\begin{equation}
\Delta_{\mathcal{T}}^2(k_0) = \frac{2\,H^2}{\pi^2 \MP^2}.
\end{equation}
Experiments do not constrain $\Delta_{\mathcal{T}}^2(k_0)$, but provide an the upper limit to the tensor-to-scalar ratio of the amplitudes of perturbations produced during inflation,
\begin{equation} \label{tensor_to_scalar_definition}
r \equiv \frac{\Delta_{\mathcal{T}}^2(k_0)}{\PkO} = \frac{2\,H^2}{\pi^2\,\MP^2\,\PkO}.
\end{equation}
The latest measurement related with tensor perturbations have been provided by the joint analysis of BICEP2/Keck Array and PLANCK data~\cite{planckBICEP1, planckBICEP2}, which constrain the tensor-to-scalar ratio $r$ at $k_0 = 0.05{\rm~Mpc}^{-1} $ as
\begin{equation} \label{r_bound}
r \lesssim 0.12, \quad \hbox{at 95\% CL}.
\end{equation}
The combination of the definition in Eq~\eqref{tensor_to_scalar_definition} with the upper bound on $r$ expressed in Eq.~\eqref{r_bound} and the scalar power spectrum $A_s$ results in an upper bound on the Hubble rate at the end of inflation~\cite{lyth1984},
\begin{equation} \label{tensor_to_scalar_upper_bound}
H \lesssim \MP\,\sqrt{\frac{0.12\,\pi^2}{2}\,\PkO} = 1.8\times 10^{13}\,{\rm~GeV}.
\end{equation}
Using the expression for $\PkO$ in Eq.~\eqref{theoretical_spectrum_full}, we obtain that the tensor-to-scalar ratio in Eq.~\eqref{tensor_to_scalar_definition}~is
\begin{equation} \label{tensor_to_scalar_As}
r = \frac{8\df_k^2}{H^2\,\MP^2}\,\frac{1}{1+2\nu_k + \omega_k} \simeq \frac{8\MP^2\,(U_\phi/U)^2}{(1+Q_k)^2\,(1+2\nu_k + \omega_k)} = \frac{16\epsilon_k}{(1+Q_k)^2\,(1+2\nu_k + \omega_k)},
\end{equation}
where in the second equality we have used Eq.~\eqref{eq_motion_slow_roll}, and in the last equality we have used Eq.~\eqref{relation_epsilon_Hdot}. In the cold inflation limit $Q_k = \nu_k = 0$ we recover the well-known condition $r = 16\epsilon_k$.

\section{Warm inflation with monomial potential} \label{Warm monomial inflation}

\subsection{Expressions for the potential and the slow-roll parameters} \label{Shape of the potential and slow-roll parameters}

Building a realistic model of warm inflation has proved to be hard in scenarios where the inflaton directly converts into light degrees of freedom, both because of large thermal corrections to the inflaton mass and because additional fields coupling to the inflaton would acquire a very large mass~\cite{berera1998, yokoyama1999}.

In facts, a Yukawa interaction $\mathcal{L}_Y = g\,\phi\,\bar{\psi}\psi$, coupling the inflaton field $\phi$ to a fermionic field $\psi$ with coupling constant $g$, would yield to a fermion mass $m_\psi \approx g\,\langle\phi\rangle$, while thermal corrections to the inflaton at temperature $T$ give $m_\phi \approx g\,T$. Both these masses are large unless the coupling $g$ is suppressed, however this option is not viable since a small value of $g$ results into inefficient dissipation to sustain a thermal bath during inflation.

For these reasons, in realistic model-building the inflaton is only indirectly coupled to radiation, firstly decaying into an intermediate hypothetical particle which then decays into radiation~\cite{berera2003, berera2005, basterogil2011, basterogil2013, basterogil2014, berera2003, basterogil2005a, basterogil2005b, moss2006, matsuda2010, basterogil2016}. Explicit formulae for the dissipation coefficient have been obtained using different approaches like particle production~\cite{abbott1982, morikawa1984}, linear response theory~\cite{hosoya1984}, and the Schwinger-Keldeysh approach~\cite{berera1998, berera2000, berera2003, moss2006}. In this latter model, dissipation is achieved in supersymmetric theories via the renormalizable superpotential
\begin{equation} \label{susy_lagrangian}
W = f(\Phi) + \frac{g}{2}\,\Phi\,X^2 + \frac{h}{2}\,X\,Y^2,
\end{equation}
where $X$, $Y$, and $\Phi$ are chiral multiplets, the inflaton field is the scalar component $\phi = \sqrt{2}\,\langle\Phi\rangle$, and the scalar potential is $U(\phi) = |f'(\phi)|^2$. During inflation, both the boson and fermion components of the $X$ field gain a large mass $m_X \approx g\,\phi/\sqrt{2}$, while the $Y$ field remains mass-less at tree-level. The $X$ field is thus an intermediary field which decays into radiation thanks to the last term in Eq.~\eqref{susy_lagrangian}~\cite{moss2006, basterogil2011, basterogil2013}, with a rate
\begin{equation} \label{gamma_susy}
\Gamma = \Gamma(T,\phi_k) = C_X\,\frac{T^3}{\phi_k^2},
\end{equation}
where $C_X$ is a dimension-less constant depending on the parameters of the supersymmetric theory. Following Eq.~\eqref{gamma_susy}, we tentatively postulate that the dissipation term is a function of both temperature and the field configuration, as
\begin{equation} \label{dissipation_monomial_c}
\Gamma = \Gamma_0\,\MP\,\left(\frac{\phi}{\MP}\right)^a\,\left(\frac{T}{\MP}\right)^b,
\end{equation}
where $\Gamma_0$ is a dimension-less constant and we introduced the constant exponents $a$~and~$b$. We show below that the temperature dependence is redundant, because of the relation between $T$ and $U$ in Eq.~\eqref{relation_T_dotphi}.

We assume that, during the slow-roll regime, the inflaton field evolves under the influence of a potential $U$ with a monomial dependence on $\phi_k$,
\begin{equation} \label{potential_monomial}
U(\phi) = \lambda\,\MP^4\,\left(\frac{\phi}{\MP}\right)^p.
\end{equation}
The inflaton potential $U(\phi)$ must be approximately flat in order to successfully explain the anisotropies observed in the CMBR~\cite{guth1982, hawking1982, starobinsky1982, bardeen1983, steinhardt1984}, with the self-coupling constant satisfying $\lambda~\lesssim~10^{-6}$~\cite{adams1991}. With this definition, the first slow-roll parameter is
\begin{equation}
\epsilon_k = \frac{p^2}{2}\,\left(\frac{\MP}{\phi_k}\right)^2,
\end{equation}
so that the temperature during the slow-roll regime, Eq.~\eqref{relation_T_dotphi}, is
\begin{equation}
\frac{T}{\MP} \simeq \sqrt[4]{\frac{\lambda\,p^2}{4\alpha}\,\frac{Q_k}{(1+Q_k)^2}\,\left(\frac{\phi}{\MP}\right)^{p-2}} = \begin{cases}
\sqrt[4]{\frac{\lambda\,p^2\,Q_k}{4\alpha}\,\left(\frac{\phi_k}{\MP}\right)^{p-2}}, & \hbox{for $Q_k \ll 1$},\\
\sqrt[4]{\frac{\lambda\,p^2}{4\alpha\,Q_k}\,\left(\frac{\phi_k}{\MP}\right)^{p-2}}, & \hbox{for $Q_k \gg 1$}.
\end{cases}
\end{equation}
Since the ratio between the dissipation term and the Hubble rate is
$$Q_k = \frac{\Gamma}{3H} = \frac{\Gamma_0}{\sqrt{3\lambda}}\,\left(\frac{\phi}{\MP}\right)^{a-p/2}\,\left(\frac{T}{\MP}\right)^b,$$
the temperature depends on the field configuration $\phi_k$ as
\begin{equation}
\frac{T}{\MP} = \begin{cases}
\left[\frac{\sqrt{3\lambda}\,p^2\,\Gamma_0}{12\alpha}\,\left(\frac{\phi_k}{\MP}\right)^{\frac{p}{2}+a-2}\right]^{\frac{1}{4-b}}, & \hbox{for $Q_k \ll 1$},\\
\left[\frac{\sqrt{3\lambda^3}\,p^2}{4\alpha\,\Gamma_0}\,\left(\frac{\phi_k}{\MP}\right)^{\frac{3p}{2}-a-2}\right]^{\frac{1}{4+b}}, & \hbox{for $Q_k \gg 1$}.
\end{cases}
\end{equation}
In the model proposed, temperature is thus dependent on $\phi_k$ during the slow-roll regime. An equivalent but simpler parametrization with respect to Eq.~\eqref{dissipation_monomial_c}, valid during slow-roll, is to assume that the dissipative term depends on $\phi_k$ only as
\begin{equation} \label{dissipation_monomial}
\Gamma = \sqrt{3\lambda}\,\gamma\,\MP\,\left(\frac{\phi}{\MP}\right)^{q/2}.
\end{equation}
Here, $\gamma$ parametrizes the strength of the dissipative term and $q$ is a constant coefficient, which are related to the parameters $\Gamma_0$, $a$, and $b$ by
\begin{eqnarray}
\gamma &=& \begin{cases}
\left[(3\lambda)^{b-2}\,\left(\frac{p^2}{12\alpha}\right)^b\,\Gamma_0^4\right]^{\frac{1}{4-b}}, & \hbox{for $Q_k \ll 1$},\\
\left[(3\lambda)^{b-2}\,\left(\frac{p^2}{12\alpha}\right)^b\,\Gamma_0^4\right]^{\frac{1}{4+b}}, & \hbox{for $Q_k \gg 1$},
\end{cases}\\
q &=& \begin{cases}
\frac{b\,p-4b+8a}{4-b}, & \hbox{for $Q_k \ll 1$},\\
\frac{3b\,p-4b+8a}{4+b}, & \hbox{for $Q_k \gg 1$}.
\end{cases}
\end{eqnarray}
We have chosen the constants in front of the $\phi$ field in Eq.~\eqref{dissipation_monomial} so that, with these values of the potential and the dissipation term, the parameter $Q$ during slow-roll reads
\begin{equation} \label{effectiveness_sl}
Q \simeq \frac{\Gamma\,\MP}{\sqrt{3U}} = \gamma\,\left(\frac{\phi}{\MP}\right)^{\frac{q-p}{2}}.
\end{equation}
For simplicity, in the following we restrict ourself to the case $q > p$ only, so that the parameter $Q$ is always proportional to some positive power of the field configuration $\phi$.


Using the potential in Eq.~\eqref{potential_monomial} and the dissipation term in Eq.~\eqref{dissipation_monomial}, the slow-roll parameters in Eq.~\eqref{slow_roll_parameters} read
\begin{eqnarray}\label{slow_roll_specific}
\epsilon_k &=& \frac{p^2}{2}\,\left(\frac{\MP}{\phi}\right)^2,\\
\eta_k &=& p\,(p-1)\,\left(\frac{\MP}{\phi}\right)^2 = 2\frac{p-1}{p}\epsilon_k,\\
\beta_k &=& \frac{p\,q}{2}\,\left(\frac{\MP}{\phi}\right)^2 = \frac{q}{p}\,\epsilon_k.
\end{eqnarray}
The slow-roll regime ends when the field $\phi$ reaches a value $\pe$ for which one of the conditions in Eq.~\eqref{slow_roll} is no longer satisfied, that is either
\begin{equation} \label{slow_roll_end}
\ee = 1+\Qe,\,\hbox{or} \,\, |\eta_{\rm end}| = 1+\Qe, \,\hbox{or}\,\, |\beta_{\rm end}| = 1+\Qe.
\end{equation}
Solving for the value of the inflaton field at the end of inflation $\pe$, we find
\begin{equation} \label{phi_end}
\pe = \begin{cases}
\MP\,\left(\frac{p\,\zeta}{2\gamma}\right)^{\frac{2}{4+q-p}}, & \hbox{for $\Qe \gg 1$},\\
\frac{\sqrt{2}p}{2}\,\MP, & \hbox{for $\Qe \ll 1$},
\end{cases}
\end{equation}
where $\zeta = \min(p,q)$, and
\begin{equation} \label{Q_end}
\Qe = \begin{cases}
\frac{p\,\zeta}{2}\,\left(\frac{2\gamma}{p\,\zeta}\right)^{\frac{4}{4+q-p}}, & \hbox{for $\gamma \gg 1$},\\
\gamma\,\left(\frac{\sqrt{2}p}{2}\right)^{\frac{q-p}{2}}, & \hbox{for $\gamma \ll 1$}.
\end{cases}
\end{equation}
When the combination $4+q-p$ is of order one or larger, the inflaton field at the end of inflation is of the order of $\pe \gtrsim \MP$, while for $4+q \ll p$ the field configuration is small compared with the Planck scale, $\pe \ll \MP$. However, the value of the field $\pe$ alone is not sufficient to describe different inflationary models. A quantity often used to label inflationary models is the scalar field excursion
$$\Delta \phi\equiv \phi_k - \pe,$$
which distinguishes between ``large-field'' inflation models for which $\Delta \phi \gg \MP$ and ``small-field'' models where $\Delta \phi \ll \MP$~\cite{lyth2008}. One simple and elegant large-field model in cold inflation is the $\phi^2$ potential~\cite{piran1985, belinsky1985}, which is the benchmark of the chaotic inflation model~\cite{linde1983}, as recently reviewed in Ref.~\cite{creminelli2014}. In the opposite limit $Q \gg 1$ and $q > p$, the value of the field excursion depends on the parameters $\gamma$, $p$, and $q$. Small-field inflation models are compelling when building theories in which quantum gravity correction at the Planck scale are avoided. We discuss these details more in depth in Sec.~\ref{application}, where we apply the definition of $\Delta \phi$ to monomial warm inflation.

\subsection{Number of e-folds} \label{Number of e-folds}

During inflation, the scale factor grows from an initial value $a_k$ to the value $a_{\rm end}$ when one of the slow-roll conditions in Eq.~\eqref{slow_roll_end} is violated. We look for the minimal amount of inflation needed to solve the flatness problem. Sufficient inflation requires
\begin{equation}\label{number_efolds_theory}
N_e \equiv \log\frac{a_{\rm end}}{a_k} \geq \log\frac{a_0}{a_{\rm end}} = \frac{1}{4}\,\log \frac{\rho_r(a_{\rm end})}{\rho_r(a_0)},
\end{equation}
where $a_0$ is the scale factor at present time and $\rho_r(a_0)\sim (10^{-4}{\rm~eV})^4$ is the present energy density in radiation. When inflation takes place at energies $\rho_r(a_{\rm end}) \sim (10^{14}{\rm~GeV})^4$, we find
\begin{equation} \label{number_efolds}
N_e \gtrsim 60,
\end{equation}
thus if inflation takes place around the $10^{14}{\rm~GeV}$ scale, it should last for at least 60 e-folds. However, in some models in which the energy scale of inflation can be as low as $\sim (1{\rm~TeV})^4$, the bound in Eq.~\eqref{number_efolds} can be pushed as low as $N_e \gtrsim 35$~\cite{lyth1999}.

Using the definition of the Hubble rate,the definition for the number of e-folds required during inflation in Eq.~\eqref{number_efolds_theory} can be written as
\begin{equation} \label{number_efoldings_th}
N_e \equiv \log\frac{a_{\rm end}}{a_k} = \int_t^{t_{\rm end}}\, H\,dt = \int_{\phi_k}^{\phi_{\rm end}}\, H\,\frac{d\phi}{\df} \simeq \frac{1}{\MP^2}\,\int_{\pe}^{\phi_k}\, \frac{U}{U_\phi}\,\left(1+Q\right)\,d\phi,
\end{equation}
where, during the inflationary stage, the value of the inflaton field decreases from the initial value $\phi_k$ to the value at the end of inflation $\pe$ defined via Eq.~(\ref{phi_end}). The expression for $N_e$ in Eq.~\eqref{number_efoldings_th} provides a relation between $\phi_k$ and $\pe$,
\begin{equation} \label{number_efoldings1}
N_e \simeq \frac{2}{4+q-p}\,\frac{\gamma}{p}\,\left[\left(\frac{\phi_k}{\MP}\right)^{\frac{4+q-p}{2}} - \left(\frac{\pe}{\MP}\right)^{\frac{4+q-p}{2}}\right] + \frac{1}{2p}\left[\left(\frac{\phi_k}{\MP}\right)^2-\left(\frac{\pe}{\MP}\right)^2\right],
\end{equation}
Using the result for $\pe$ in Eq.~\eqref{phi_end}, the relation in Eq.~\eqref{number_efoldings_th} reads
\begin{equation} \label{number_efoldings2}
N_e \simeq \begin{cases}
\frac{2\gamma}{p(4+q-p)}\,\left(\frac{\phi_k}{\MP}\right)^{\frac{4+q-p}{2}} - \frac{\zeta}{4+q-p}, & \hbox{for $\gamma \gg 1$},\\
\frac{1}{2p}\left(\frac{\phi_k}{\MP}\right)^2-\frac{p}{4}, & \hbox{for $\gamma \ll 1$}.
\end{cases}
\end{equation}
The relation in Eq.~\eqref{number_efoldings2} can be inverted to give the value of the inflaton field at horizon crossing $\phi_k$ as a function of the number of e-folds $N_e$,
\begin{equation} \label{relation_phi_Ne}
\frac{\phi_k}{\MP} \simeq \begin{cases}
\left\{\frac{p[(4+q-p)N_e+\zeta]}{2\gamma}\right\}^{\frac{2}{4+q-p}}, & \hbox{for $\gamma \gg 1$},\\
\sqrt{\frac{p(4N_e+p)}{2}}, & \hbox{for $\gamma \ll 1$}.
\end{cases}
\end{equation}
Given the relation in Eq.~\eqref{relation_phi_Ne}, the expression for the first slow-roll parameter at horizon crossing is obtained from Eq.~\eqref{slow_roll_specific}, as
\begin{equation} \label{epsilon_Ne}
\epsilon_k \simeq \begin{cases}
\frac{p^2}{2}\,\left[\frac{2\gamma}{p(4+q-p)N_e+\zeta]}\right]^{\frac{4}{4+q-p}}, & \hbox{for $\gamma \gg 1$},\\
\frac{p}{4\,N_e+1}, & \hbox{for $\gamma \ll 1$}.
\end{cases}
\end{equation}
Similarly, the expression for the dissipation parameter $Q_k$ is obtained by inserting $\phi_k$ in Eq.~\eqref{relation_phi_Ne} into Eq.~\eqref{effectiveness_sl},
\begin{equation} \label{Q_Ne}
Q_k \simeq \begin{cases}
\gamma\,\left\{\frac{2\gamma}{p[(4+q-p)N_e+\zeta]}\right\}^{\frac{-q+p}{4+q-p}}, & \hbox{for $\gamma \gg 1$},\\
\gamma\,\left[\frac{p(4N_e+p)}{2}\right]^\frac{q-p}{4}, & \hbox{for $\gamma \ll 1$}.
\end{cases}
\end{equation}

\subsection{Strong limit of warm inflation} \label{Strong limit of warm inflation}

From here to the end of the paper, we restrict ourselves to considering monomial warm inflation in the strong dissipation limit, $Q \gg 1$. In this scenario, inflation ends when one of the slow-roll parameters reaches a value
$$\Qe = \frac{p\,\zeta}{2}\,\left(\frac{2\gamma}{p\,\zeta}\right)^{\frac{4}{4+q-p}}.$$

In the strong limit of warm inflation, the number of e-folds is given by the first line of Eq.~\eqref{number_efoldings2}, a relation which can be inverted to find $\phi_k$ at horizon crossing as in the first line of Eq.~\eqref{relation_phi_Ne}. The corresponding values of the first slow-roll parameter $\epsilon_k$ and the dissipation parameter $Q_k$ at scale $k$ are respectively given by the first lines of Eq.~\eqref{epsilon_Ne} and Eq.~\eqref{Q_Ne}.
The power spectrum in the strong limit of warm inflation is given by Eq.~\eqref{theoretical_spectrum_full} in the limit $Q_k~\gg~1$,
\begin{equation} \label{scalar_spectrum_strong}
\PkO = \frac{H^4}{4\pi^2\,\df_k^2}\,\omega_k = \frac{H^{5/2}\,T\,\Gamma^{1/2}}{4\pi^{3/2}\,\df_k^2}= \frac{3H^3\,T\,Q_k^{1/2}}{4\pi^{3/2}\,\df_k^2},
\end{equation}
a result previously obtained in the literature, except for the numerical factor in front~\cite{taylor2000, basterogil2005b, moss2008, visinelli2011}. Similarly, the strong dissipation limit of the tensor-to scalar ratio is obtained starting from Eq.~\eqref{tensor_to_scalar_As}
\begin{equation} \label{tensor_to_scalar_strong}
r \simeq \frac{16\epsilon_k}{Q_k^2\,\omega_k} = \frac{H}{T}\,\frac{16\epsilon_k}{\sqrt{3\pi}\,Q_k^{5/2}}.
\end{equation}
Notice that, for a given value of $\epsilon_k$, the corresponding tensor-to-scalar ratio in warm inflation is smaller than the corresponding result in cold inflation both because of thermal effects $H < T$ and the dissipation term $Q_k \gg 1$. Finally, the strong dissipation limit of the scalar spectral tilt has been obtained in the first line of Eq.~\eqref{spectral_tilt_warm_strong1}, which can be re-expressed in terms of $\epsilon_k$ only by using the identities in Eq.~\eqref{slow_roll_specific} valid for monomial inflation,
\begin{equation} \label{index_strong}
n_s-1 = -\frac{3\left(4+3q-p\right)}{4\,p\,Q_k}\,\epsilon_k.
\end{equation}

We now express the observables $A_s$, $n_s$, and $r$ in terms of the parameters of monomial warm inflation $p$, $q$, $\lambda$, and the number of e-folds $N_e$ only. For this, we first eliminate the terms $\df_k$, $H$, and $T$ in the expressions for the observables by using the following useful identities valid in the strong dissipation limit,
\begin{equation} \label{gamma_gg_1_phi_T}
\df_k^2 \simeq \frac{2\,U\,\epsilon_k}{3Q_k^2},\quad H^2 \simeq \frac{U}{3\MP^2} \quad \rho_r \simeq \frac{U\,\epsilon_k}{2Q_k}, \quad T \simeq \sqrt[4]{\frac{U\,\epsilon_k}{2\alpha\, Q_k}}.
\end{equation}
%
Inserting these relations in Eqs.~\eqref{scalar_spectrum_strong},~\eqref{tensor_to_scalar_strong}, and~\eqref{index_strong}, we obtain the values of the observable quantities in terms of the field configuration $\phi_k$ only,
\begin{eqnarray}
\PkO &=& \frac{1}{8\pi}\,\sqrt[4]{\frac{9}{2\pi^2\,\alpha}}\left(\frac{Q^3\,U}{\MP^4\epsilon_k}\right)^{3/4} = \frac{1}{8\pi}\,\sqrt[4]{\frac{36\,\gamma^9\,\lambda^3}{\pi^2\,\alpha\,p^6}}\,\left(\frac{\phi_k}{\MP}\right)^{\frac{3(4+3q-p)}{8}}, \label{relation_As_phi_strong}\\
n_s-1 &=& -\frac{3\left(4+3q-p\right)}{4\,p\,Q_k}\,\epsilon_k = -\frac{3p\left(4+3q-p\right)}{8\,\gamma}\,\left(\frac{\phi_k}{\MP}\right)^{\frac{4+q-p}{2}}, \label{relation_ns_gamma}\\
r &=& \frac{16}{3} \sqrt[4]{\frac{2\alpha\,U\,\epsilon_k^3}{\pi^2\,\MP^4\,Q_k^9}} = \frac{16}{3} \sqrt[4]{\frac{\alpha\,\lambda\,p^6}{4\pi^2\,\gamma^9}}\,\left(\frac{\phi_k}{\MP}\right)^{-\frac{12+9q-11p}{8}}, \label{r_phi_strong}
\end{eqnarray}
where the last equalities in the expression above have been obtained using Eq.~\eqref{potential_monomial} for $U$, Eq.~\eqref{effectiveness_sl} for $Q_k$, and Eq.~\eqref{slow_roll_specific} for $\epsilon_k$, since all of these equation depend on $\phi_k$ only. 

\subsection{Application to some popular inflation models} \label{application}

We now turn our attention to three inflation models, namely the axion monodromy model~\cite{silverstein2008, mcallister2010}, where the inflaton potential is either $U(\phi)\sim \phi^{2/3}$ or $U(\phi)\sim \phi$, the chaotic inflation model~\cite{linde1983} with $U(\phi)\sim \phi^2$, and a $U(\phi)\sim \phi^3$ model. These choices respectively correspond to setting $p = 2/3$, $p=1$, $p=2$, or $p=3$ in Eq.~\eqref{potential_monomial}.

\begin{figure}[h!]
\begin{center}
\includegraphics[height=10cm]{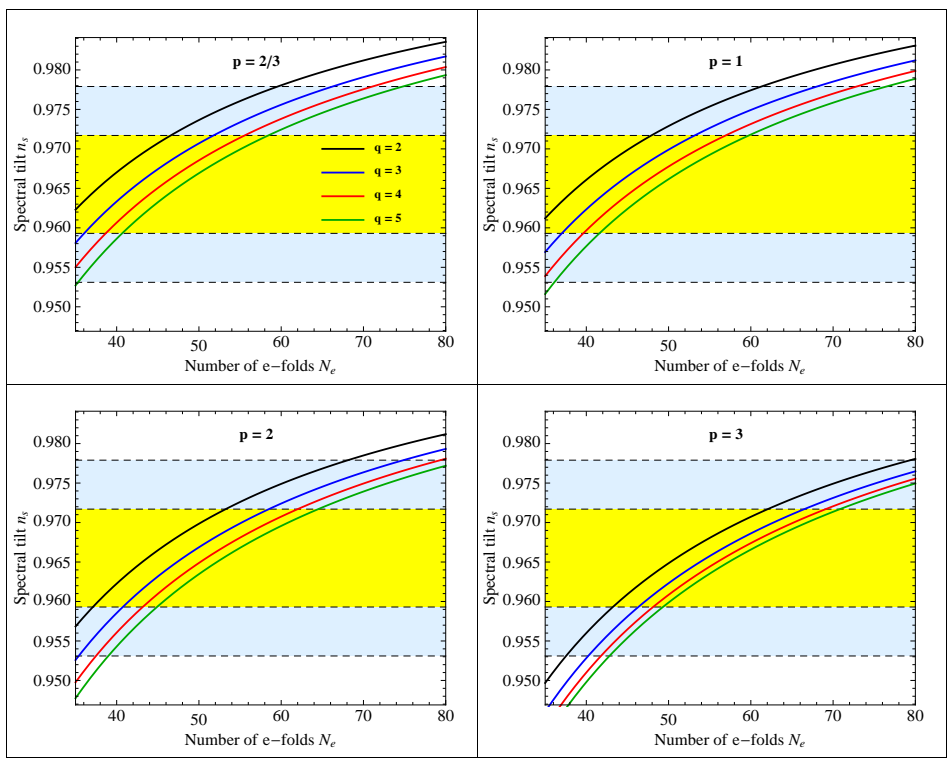}
\caption{The scalar spectral tilt $n_s$ as a function of the number of e-folds $N_e$, obtained from Eq.~\eqref{relation_ns_gamma1}. We have considered the models with the potential of index $p = 2/3$, $p=1$, $p=2$, and $p=3$. For each model, we show results for $q = 2$ (black), $q = 3$ (blue), $q=4$ (red), and $q=5$ (green). The yellow band shows the 68\% confidence region on $n_s$ as measured by PLANCK , with the uncertainties given by Eq.~\eqref{spectral_tilt_measured}. The light blue band shows the 95\% confidence region.}
\label{plot_ns}
\end{center}
\end{figure}

Using the expressions for $\epsilon_k$ and $Q_k$ in the first lines of Eqs.~\eqref{epsilon_Ne} and~\eqref{Q_Ne}, we replace the field $\phi_k$ with a function containing $N_e$ so that
\begin{equation} \label{relation_ns_gamma1}
n_s = 1 - \frac{3p\left(4+3q-p\right)}{4\,\left[p\,(4+q-p)\,N_e-\zeta\right]}.
\end{equation}
In Fig.~\ref{plot_ns} we show the value of $n_s$ as a function of the number of e-folds $N_e$, obtained from Eq.~\eqref{relation_ns_gamma1}. In Fig.~\ref{plot_ns}, we have set the values $q = 2$ (black), $q = 3$ (blue), $q=4$ (red), and $q=5$ (green). In the figure, the width of the yellow stripe corresponds to the 68\% CL of $n_s$ and the light blue regions define the 95\% CL, as obtained from Eq.~\eqref{spectral_tilt_measured}. For smaller values of $p$, corresponding to shallower potentials, lower values of $N_e$ are favored.

\begin{figure}[h!]
\begin{center}
\includegraphics[height=10cm]{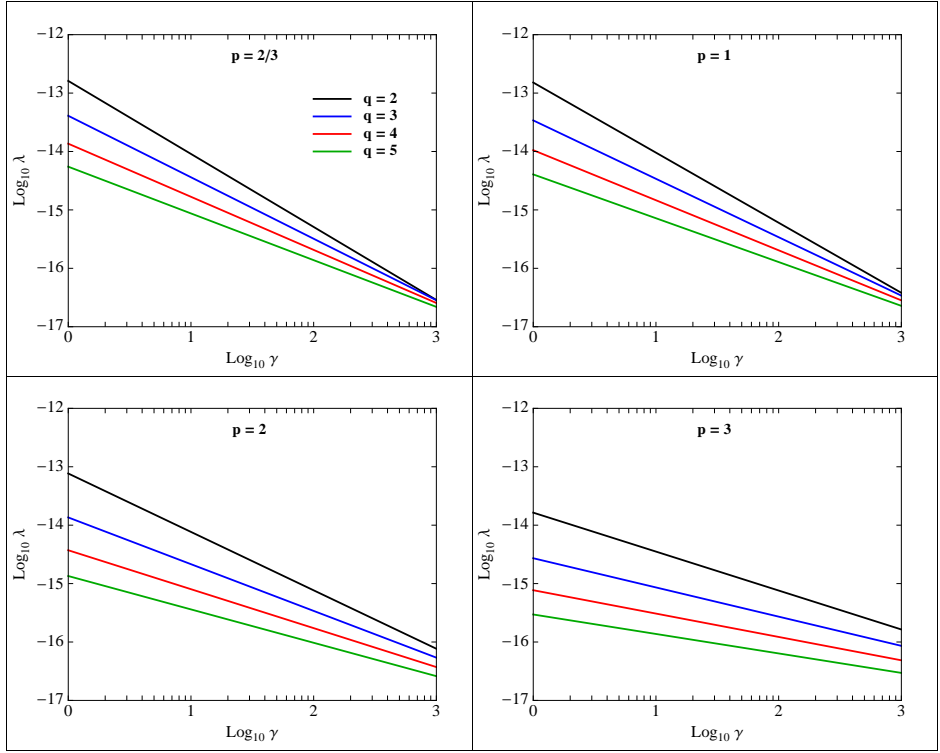}
\caption{The coupling $\lambda$ as a function of the dissipation strength $\gamma$, obtained from Eq.~\eqref{relation_lambda_gamma}. We have considered the models with the potential $p = 2/3$, $p=1$, $p=2$, and $p=3$. For each model, we show results for $q = 2$ (black), $q = 3$ (blue), $q=4$ (red), and $q=5$ (green).}
\label{fig_lambda_gamma}
\end{center}
\end{figure}

Thanks to the expression in Eq.~\eqref{relation_ns_gamma1}, we trade $N_e$ for $\gamma$ in the first line of Eq.~\eqref{relation_phi_Ne}, so that the field $\phi_k$ is given in terms of $n_s$ as
\begin{equation} \label{relation_phi_ns}
\frac{\phi_k}{\MP} = \left[\frac{3p\,(4+3q-p)}{8\gamma\,(1-n_s)}\right]^{\frac{2}{4+q-p}}.
\end{equation}
Inserting Eq.~\eqref{relation_phi_ns} into Eq.~\eqref{relation_As_phi_strong}, we find the expressions for the power spectrum
\begin{equation}
\PkO = \frac{1}{8\pi}\,\sqrt[4]{\frac{36\,\gamma^9\,\lambda^3}{\pi^2\,\alpha\,p^6}}\,\left[\frac{3p\,(4+3q-p)}{8\gamma\,(1-n_s)}\right]^{\frac{3(4+3q-p)}{4(4+q-p)}} \label{relation_As_phi_strong1}.
\end{equation}
Eq.~\eqref{relation_As_phi_strong1} gives a relationship between $\lambda$ and $\gamma$ that depends on the observables $\PkO$ and $n_s$ and on the parameters $q$ and $p$. Since $A_s$ and $n_s$ have been measured from experiments, we invert Eq.~\eqref{relation_As_phi_strong1} to yield a relationship between $\lambda$ and $\gamma$,
\begin{equation} \label{relation_lambda_gamma}
\lambda = \frac{8\pi^2\,p^2}{3\gamma^3}\,\sqrt[3]{\frac{\pi^2\,g_*(T)}{5}\,A_s^4}\,\left[\frac{8\gamma\,(1-n_s)}{3p\left(4+3q-p\right)}\right]^\frac{4+3q-p}{4 + q - p},
\end{equation}
where we have set $\alpha = \pi^2\,g_*(T)/30$. Figure~\ref{fig_lambda_gamma} shows the value of the parameter $\lambda$ as a function of the dissipation strength $\gamma$, obtained from the expression in Eq.~\eqref{relation_lambda_gamma}, setting the value of the power spectrum $A_s$ equal to the observed value by PLANCK in Eq.~\eqref{constraint_power_spectrum} and the value of $n_s$ given in Eq.~\eqref{spectral_tilt_measured}. In Fig.~\ref{fig_lambda_gamma}, we used the values $q = 2$ (black), $q = 3$ (blue), $q=4$ (red), and $q=5$ (green). The uncertainty associated with each yellow band depends on the uncertainties on $A_s$ and $n_s$, given by 
\begin{equation} \label{uncertainty_As}
\Delta \lambda^2 = \left|\frac{\partial \lambda}{\partial A_s}\right|^2\,\Delta A_s^2 +  \left|\frac{\partial \lambda}{\partial n_s}\right|^2\,\Delta n_s^2.
\end{equation}
The first term in Eq.~\eqref{uncertainty_As} is much smaller than the second one, so that most of the uncertainty on $\lambda$ comes from the measurement of $n_s$.
Taking for example $\gamma = 10^3$, we estimate the order of magnitude of some important quantities in the theory, as reported in Table~\ref{Table_magnitude}. The mass of the the inflaton field $m$ has been obtained assuming a quadratic potential $U(\phi) = m^2\,\phi^2/2$, or $m^2 = \lambda\,\MP^2$.
\begin{table}[h!]
\caption{The order of magnitude of some relevant quantities in monomial warm inflation, for $\gamma = 10^3$.}
\begin{center}
\label{Table_magnitude}
\begin{tabular}{|l|c|}
\hline
Coupling $\lambda$ & $10^{-17}$\\
Tensor-to-scalar ratio $r$ & $10^{-9}$\\
Inflaton mass $m$ & $10^{10}{\rm~GeV}$\\
Potential height $U^{1/4}$ & $10^{14}{\rm~GeV}$\\
Hubble rate $H$ & $10^{10}{\rm~GeV}$\\
Dissipation parameter $\Gamma$ & $10^{13}{\rm~GeV}$\\
Radiation temperature $T$ & $10^{13}{\rm~GeV}$\\
Field excursion $\Delta \phi$ & $0.01\,\MP$\\
\hline
\end{tabular}
\end{center}
\end{table}

\begin{figure}[t!]
\begin{center}
\includegraphics[height=10cm]{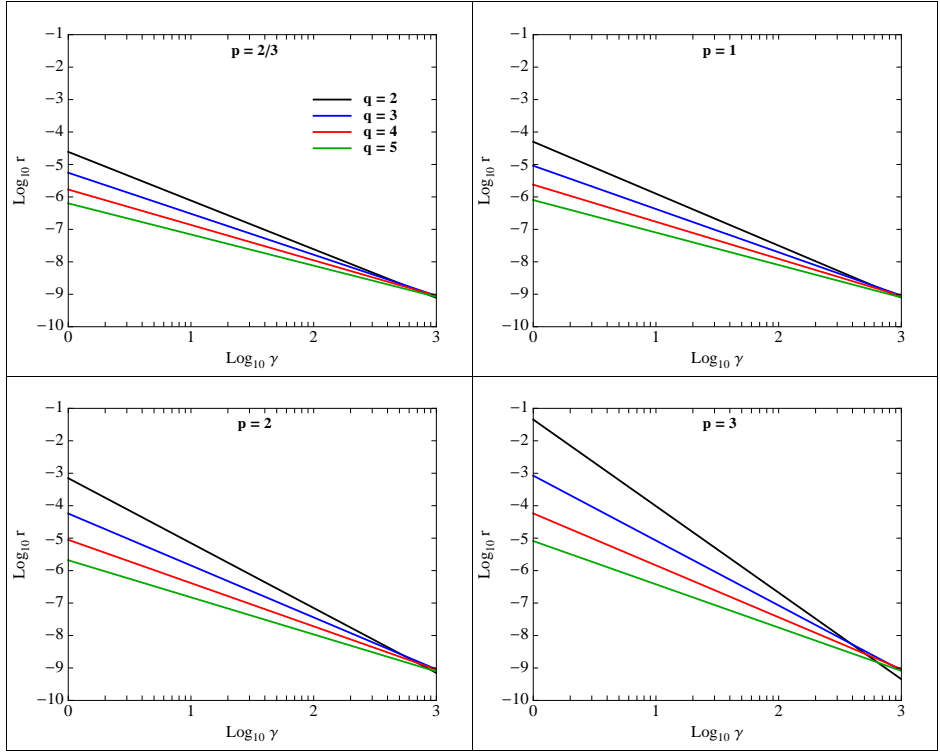}
\caption{The tensor-to-scalar ratio $r$ as a function of the dissipation strength $\gamma$, obtained from Eq.~\eqref{r_phi_strong1}. We show results for $p = 2/3$, $p=1$, $p=2$, and $p=3$. For each model, we plot the lines for $q = 2$ (black), $q = 3$ (blue), $q=4$ (red), and $q=5$ (green).}
\label{r_Ne}
\end{center}
\end{figure}

The expression for the tensor-to-scalar ratio in Eq.~\eqref{r_phi_strong} can also be given in terms of $\gamma$ by replacing $\phi_k$ with the value in Eq.~\eqref{relation_phi_ns}, to obtain
\begin{equation}
r = \frac{16}{3} \sqrt[4]{\frac{\alpha\,\lambda\,p^6}{4\pi^2\,\gamma^9}}\,\left[\frac{3p\,(4+3q-p)}{8\gamma\,(1-n_s)}\right]^{-\frac{12+9q-11p}{4(4+q-p)}} \label{r_phi_strong1}.
\end{equation}
In Fig.~\ref{r_Ne}, we show the tensor-to-scalar ratio $r$ as a function of the dissipation strength $\gamma$, obtained from Eq.~\eqref{r_phi_strong1}. For any value of $\gamma$, the values of the tensor-to-scalar ratio are much smaller than the bound in Eq.~\eqref{r_bound}, set by the PLANCK mission. In particular, for $\gamma = 10^3$ we obtain a tensor-to-scalar ratio of the order of $r \sim 10^{-9}$.
As discussed in Sec.~\ref{Shape of the potential and slow-roll parameters}, in cold inflation such a low value of $r$ is typical of small-field inflationary models~\cite{lyth2008}.

\begin{figure}[b!]
\begin{center}
\includegraphics[height=10cm]{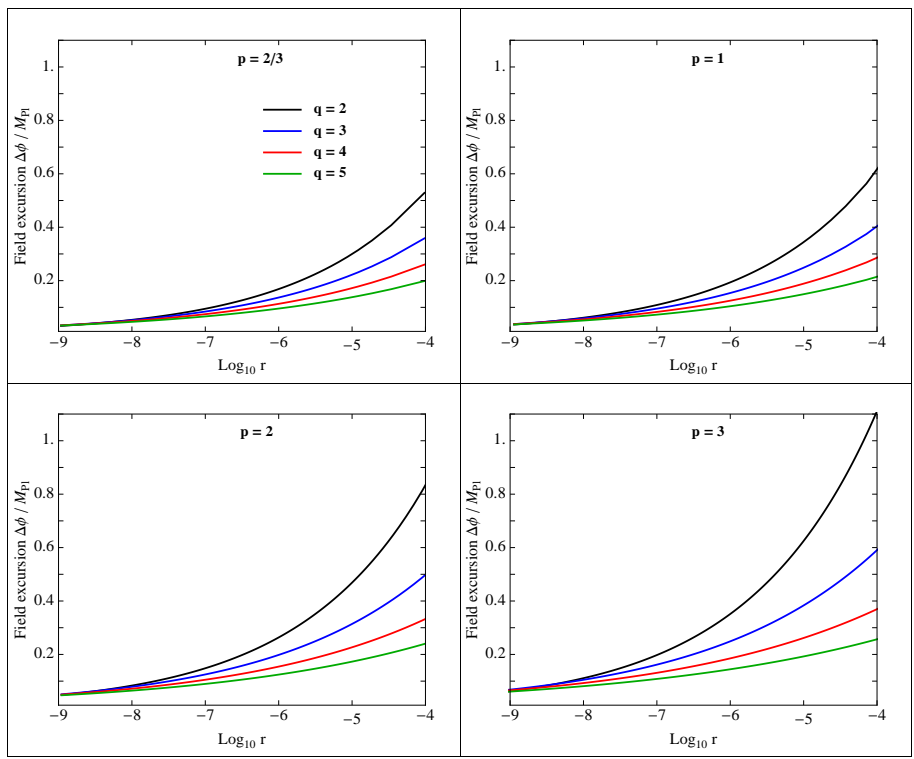}
\caption{The field excursion $\Delta \phi$ as a function of the tensor-to-scalar ratio $r$, obtained from Eq.~\eqref{lythbound1} for $q = 2$ (black), $q = 3$ (blue), $q=4$ (red), and $q=5$ (green).}
\label{field_excursion}
\end{center}
\end{figure}

We investigate the nature of monomial warm inflation by deriving the relation between the number of e-folds and the field excursion
$$\Delta \phi\equiv \phi_k - \pe.$$
For this, we first use the Friedmann equation $U = 3\MP^2\,H^2$ to obtain the derivative of $U$ with respect to the inflaton field $\phi$,
\begin{equation}
\frac{U_\phi}{U} = \frac{2H_\phi}{H}.
\end{equation}
The expression for the first slow-roll parameter $\epsilon_k$ is then
\begin{equation}
\epsilon_k = 2\MP^2\,\left(\frac{H_\phi}{H}\right)^2 = 2\MP^2\,\left(\frac{\dot{H}}{H\,\df}\right)^2,
\end{equation}
which can be expressed in terms of the number of e-folds in Eq.~\eqref{number_efoldings_th} and the expression for $\epsilon_k$ in Eq.~\eqref{slow_roll_specific} as
\begin{equation} \label{dphi_over_dN}
\frac{d \phi}{dN_e} = \MP\,\frac{\dot{H}}{H}\,\left(\frac{dN_e}{dt}\right)^{-1}\,\left(\frac{2}{\epsilon_k}\right)^{1/2} = \MP\,\frac{\sqrt{2\epsilon_k}}{1+Q_k}.
\end{equation}
In the limit $Q_k \ll 1$ we obtain the result in the cold theory of inflation, while the strong-dissipation limit $Q_k \gg 1$, which is the one relevant in this discussion, gives
\begin{equation} \label{dphi_over_dN1}
\frac{d \phi}{dN_e} = \MP\,\frac{\sqrt{2\epsilon_k}}{Q_k}.
\end{equation}
Combining this latter equality with the consistency relation in Eq.~\eqref{tensor_to_scalar_strong}, we finally obtain the expression for the field excursion in terms of the number of e-folds,
\begin{equation} \label{lythbound1}
\Delta\phi = \frac{\MP}{4}\,\sqrt[4]{6\pi\,Q_k\,\left(\frac{r\,T}{H}\right)^2}\,N_e.
\end{equation}
The number of e-folds $N_e$ required to inflate between the modes with CMB multipoles $2 \leq l \leq 100$ at horizon crossing is $N_e \approx 4$, while sufficient inflation requires $N_e > 35$ as discussed earlier. This yields to a lower bound on the scalar field excursion as a function of $r$, known in the literature as the ``Lyth bound''~\cite{lyth1997}. In cold inflation models, the Lyth bound constrain large field theories to predict relatively large (observable) values of the tensor-to-scalar ratio. Here, Eq.~\eqref{lythbound1} is thus the analogous formulation of the Lyth bound in warm inflation. We write $\Delta\phi$ as a fun`ction of the dissipation strength $\gamma$ only, by eliminating $N_e$ using Eq.~\eqref{relation_ns_gamma1} and the coupling $\lambda$ using Eq.~\eqref{relation_As_phi_strong1}. Results are shown in Fig.~\ref{field_excursion} for the value of $\Delta \phi$ as a function of $r$, with the potential of index $p = 2/3$, $p=1$, $p=2$, and $p=3$.
In warm inflation, the inflaton field might achieve a large excursion $\Delta \phi \sim \MP$ with negligible value of the tensor-to-scalar ratio $r \ll 0.01$, thus evading the prediction valid in cold inflation models. As evident from Eq.~\eqref{lythbound1}, this result is possible since $r$ is multiplied by the quantities $Q_k$ and $T/H$, which in the strong limit of warm inflation are both greater than one. The possibility that the ordinary Lyth bound might be evaded in warm inflation has been also discussed in Refs.~\cite{cai2011, bartrum2013}, and can be used to experimentally distinguish between warm and cold inflationary models, while the possibility of eluding the Lyth bound in cold inflation models with polynomial potentials has been considered in Ref.~\cite{hotchkiss2012}. To sum up, warm inflation is one of the best motivated inflationary models to fit current observations. Monomial warm inflation naturally accommodates values of the tensor-to-scalar ratio which are much smaller than the current bound set by the BICEP2-PLANCK mission. For typical values of the parameters in the theory, we have obtained $r\sim 10^{-9}$. At the same time, the inflaton field evolves spanning a large value $\Delta \phi\sim 0.01\MP$, which in the cold theory of inflation is associated with a much larger tensor-to-scalar ratio than what obtained in monomial warm inflation. Using the observational value of the scalar spectral tilt $A_s$, we have obtained a value of the coupling constant $\lambda \sim 10^{-17}$, which leads to an estimate of the parameters of the theory as in Table~\ref{Table_magnitude}. The Hubble rate obtained is $H\sim 10^{10}{~\rm GeV}$, which is safely within the bound given by Eq.~\eqref{tensor_to_scalar_upper_bound} and is much smaller than $\Gamma \sim 10^{13}{\rm~GeV}$ thus justifying the strong dissipation treatment we adopted.


\begin{acknowledgments}
The author would like to thank \O{}yvind Gr\o{}n (U. Oslo) for useful comments and for finding a mistake in an earlier version of this paper.
\end{acknowledgments}

\end{document}